\documentclass[pre,preprint,showpacs,superscriptaddress]{revtex4}
\usepackage{graphicx}
\usepackage{psfrag}
\usepackage[sort&compress]{natbib}

\newcommand{\eqref}[1]{(\ref{#1})}
\newcommand{\be}{\begin{equation}}
\newcommand{\ee}{\end{equation}}
\newcommand{\te}[1]{\mbox{$ \bf #1$}}

\newcommand{\Frac}[2]{\frac{\displaystyle #1}{\displaystyle #2}}
\newcommand{\der}[2]{\Frac{\mathrm{d}#1}{\mathrm{d}#2}}
\newcommand{\pder}[2]{\frac{\displaystyle\partial#1}{\displaystyle\partial#2}}
\newcommand{\bea}{\begin{eqnarray}}
\newcommand{\eea}{\end{eqnarray}}
\newcommand{\ob}[1]{\overline{#1}}

\newcommand{\gamdot}{\mbox{$\gamma$}}

\newcommand\textfrac[2]{{\textstyle #1}/{\textstyle #2}}

\newcommand{\Ca}{C\!a}
\def\old{\cite{govindarajan_etal01}}
\def\acrcom{\cite{acrivos_comment02}}
\def\tiruseg{\cite{tirumkudulu_etal99,tirumkudulu_etal00}}
\def\segall{\cite{tirumkudulu_etal99,tirumkudulu_etal00,boote_thomas99,thomas_etal01}}
\begin{document}

 \title{The segregation instability of a sheared suspension film}
\author{Rama Govindarajan}
\email{rama@jncasr.ac.in}
\affiliation{Engineering Mechanics Unit,
Jawaharlal Nehru Centre for Advanced Scientific
Research, Jakkur, Bangalore 560 064, India}
\author{Prabhu R. Nott}
\email{prnott@chemeng.iisc.ernet.in}
\affiliation{Department of Chemical Engineering,
Indian Institute of Science, Bangalore 560 012, India}
\author{Sriram Ramaswamy\footnote{Also with JNCASR, Bangalore 560 064 India}}
\email{sriram@physics.iisc.ernet.in}
\affiliation{Centre for Condensed-Matter Theory, Department of Physics,
Indian Institute of Science, Bangalore 560 012, India}

\begin{abstract}
\baselineskip 12pt
Starting from the equations of Stokes flow and the mass conservation of particles
as determined by shear-induced diffusion, 
we derive the coupled equations for the dynamics of particle concentration 
and film thickness for the free-surface flow of a fluid film 
pulled up by a tilted wall rising from a pool of neutrally buoyant, 
non-Brownian suspension.  
We find an instability of the film with respect to axial undulations of 
film thickness and modulations of particle concentration, and the 
instability growth-rate increases as a certain combination of the two dimensionless shear
induced diffusivities (which determine the particle flux driven by concentration and 
shear rate gradients) falls below a critical value.
This reinforces the conclusions of {\it Phys. Fluids} {\bf 13} (12), p.\ 3517 (2001), suggesting an explanation of the experiments of Tirumkudulu {\it et al.}, Phys. Fluids {\bf 11}, 507-509 (1999); {\it ibid.} {\bf 12}, 1615 (2000).  In addition, we predict a ``pile-up'' instability in which perturbations that vary in the direction of the wall velocity are amplified; this instability is not driven by shear-induced migration, but is a result of the dependence of the suspension viscosity on the particle concentration.
\end{abstract}
\pacs{47.55.Kf Multiphase and particle-laden flows; 47.54.+r Pattern selection, pattern formation; 47.15.Gf Low-Reynolds-number (creeping) flows; 47.20.Gv Viscous instability}

\maketitle

\vspace*{2em}

\section{Introduction}
\label{intro}

\subsection{Background}
\label{backgrd}
Recent studies 
\segall\ have reported 
observations of axial segregation of particles in a suspension contained in a horizontal 
cylinder rotating about its axis.  While the segregation of particles in flowing 
suspensions has been widely studied in a variety of flows, there is an important 
difference in the geometry studied by 
\segall, namely, the 
presence of a free interface of the suspension: segregation was found to occur only when 
the cylinder is {\em partially} filled with suspension.
These experiments have led to attempts at theoretical analysis 
\old, further  experiments 
\cite{timberlake_morris02}, and criticism \acrcom.  As in 
earlier experiments on shear-induced migration, the suspension comprised small neutrally 
buoyant particles dispersed in fluid of viscosity much higher than that of water, so that 
the Reynolds number (based on particle size) was negligibly small, and the Peclet number 
was very large.  The former implies the irrelevance of particle and fluid inertia in 
determining the state of the system, and the latter means that Brownian motion of the 
particles is negligible.  Particle segregation in suspensions has important 
consequences in industrial processing, where it is usually important to disperse the particles 
uniformly to obtain a product of predictable quality.  It is also possible that 
segregation could be used to separate particles from the fluid, or particles 
of different sizes from each other. From a fundamental perspective, it is important to 
understand the physical or mechanical origin of segregation.

\begin{figure}
\psfrag{w}{$w$}
\psfrag{r}{$r$}
\psfrag{v}[b]{$v_w$}
\psfrag{cr}[bc]{\rule{0.15em}{0em}$R$}
\psfrag{s}[bc]{suspension pool}
\begin{center}
\includegraphics[scale=0.5]{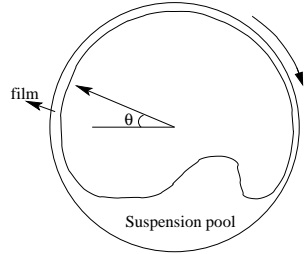}
\caption{Schematic diagram of film coating in a horizontal rotating cylinder\label{fig-schematic}}
\end{center}
\end{figure}

	In a recent paper~\old, we had proposed a model to 
explain the observations in 
\segall\ 
based on a simple and intuitive picture of the dynamics of the suspension film coating 
the wall of the cylinder.  We proposed that a balance between gravity, viscous 
shear forces, and surface tension determine the evolution of the film thickness, which 
in turn determines the shear rate in the film.  The formulation was completed by coupling this balance with the conservation equation for particle mass, in which the particle 
concentration evolves according to shear-induced diffusion \cite{leighton_acrivos87}.  Our analysis was simplified by the 
assumption that the variations of the flow variables within the cross section (i.e.\ 
in the $r$-$\theta$ plane, see Fig.~\ref{fig-schematic}) at any given axial coordinate 
are unimportant, and that segregation is driven by axial perturbations of the particle 
concentration and the film thickness.  We conducted a linear stability analysis, and 
predicted that the homogeneous suspension is unstable to small perturbations 
provided a certain discriminant, determined by the dimensionless shear-induced diffusivities and the effective viscosity of the suspension, is negative. 
An important feature of our model is the necessity of a free interface, 
which is in agreement with the observation of \tiruseg.

One of the aims of the present paper is to lay to rest once and 
for all the doubts raised -- in an {\it ad hominem} comment \acrcom\ -- about 
the validity of our approach. 
From a complete 2D analysis, we are able to show
where Eq.~(4) of \old, which is the force balance mentioned above,
comes from. Incidentally, this part of the derivation was
already presented in our rebuttal to the comment.

\subsection{Summary of results}
\label{resultsum}
Before describing the model and our calculations in detail, let us summarize 
the achievements of this paper.  We consider the coating of a plate, with coordinates in the ($x$,$z$) plane, being pulled out of a pool of suspension at constant velocity in the $x$ direction (see Fig.~\ref{fig-schematic2}).  From the Stokes equations and the particle mass balance, as determined by shear-induced diffusion
\cite{leighton_acrivos87}, we derive a pair of coupled partial differential 
equations for the film thickness and the particle concentration
as functions of time $t$ and the spatial coordinates $x$, $z$.
From these we obtain the steady-state thickness and concentration profiles 
(see Fig.~\ref{fig-base}) 
and the equations for the evolution of small perturbations about this steady state 
(see Eqs~\ref{eqn-w_lin}, \ref{eqn-phi_lin}).  The form of these equations, 
for variation solely along $z$, and ignoring surface tension, 
is identical to that presented on general physical grounds 
in \old, and criticized in \acrcom.  
Two instability mechanisms emerge from this analysis -- the  
{\em axial} instability proposed in our earlier one-dimensional treatment 
\old, and another ``pile-up'' growth along the $x$ direction (see \S\ref{sec-linear_stab}).  
As stated in our earlier work \old, 
we find that if the discriminant $E$, which depends on the dimensionless shear induced diffusivities, is below a critical value $E_c$, the growth rate
varies as $k_z^2$ for {\em small} wavenumber $k_z$ in the $z$  direction, turning over at larger wavenumbers on a scale determined by surface tension. 
There are two differences between \old\ and the present results: (i) Because the 
eigenfunctions of the growing modes all have significant $x$-dependence, the 
circumferential instability always acts, so that the growth rate is nonzero even 
for $k_z = 0$. The axial instability of \old\ rides on this background ``pile-up'' 
growth-rate.  As a result, we predict $E_c > 0$ whereas the one-dimensional analysis of
\old\ predicted $E_c=0$.  (ii) The wavenumber $q_*$ of peak growth rate varies 
as the inverse {\em fourth} root, not square root, of the surface tension. 
Figs~\ref{fig-dispnrelnsigma} and \ref{fig-dispnrelnfc} show the 
effects of changing the surface tension and the discriminant, respectively, on the   
instability growth rate as a function of $k_z$.  
These minor differences apart, the mechanism proposed in \old\
for the segregation 
instability of \segall\ remains viable.  

	The remainder of this paper is organized as follows. 
Starting from the Stokes equations and the kinematic condition at a free interface, we first derive in \S\ref{sec-full_model} the equation for the dynamics of the film thickness, making only the lubrication approximation as in almost all studies of the drainage of viscous films.  We show that this equation reduces to Eq.~(4) of \old\ (save the surface tension term) if certain simplifying assumptions are made on the variation of the film thickness in the direction of wall motion.  In \S\ref{sec-linear_stab} we conduct a linear stability analysis of this equation, coupled with the conservation equation for particle mass, and show that it yields essentially the same result as that the simple model we had proposed in \old.  We close the paper with a summary and conclusions in \S \ref{concl}.

\section{Dynamics of a suspension film with a free interface}
\label{sec-full_model}

\begin{figure}
\psfrag{q}{$\theta$}
\psfrag{r}{$r$}
\psfrag{vw}[b]{$v_w$}
\begin{center}
\includegraphics[scale=0.5]{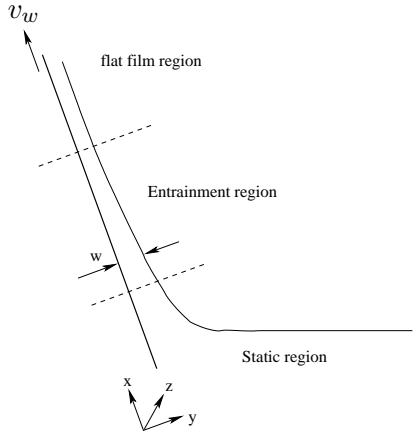}
\caption{{Coating of a vertical plate rising with velocity $v_w$ from a pool of suspension.}}
\label{fig-schematic2}
\end{center}
\end{figure}

	We consider the coating of a flat plate rising from a pool of suspension at an angle $\theta$ from the vertical, shown in Fig.~\ref{fig-schematic2}.  Several studies \cite{landau_levich42,middleman77,white_tallmadge65,spiers_etal74} have investigated the classical problem of coating of a flat vertical plate rising from a pool of liquid.  The object of all these studies was to determine the asymptotic film thickness far from the pool, as that is the primary variable of interest in industrial coating operations.  The dynamics and stability of coating flows of pure liquids also have been analyzed \cite{moffat77,hosoi_mahadevan99}, in an effort to explain instabilities and pattern formation that are observed \cite{balmer70,karweit_corrsin75,melo93,thoroddsen_mahadevan97}.  In this study, we consider the dynamics of the coating flow of a suspension whose viscosity is a function of the particle concentration.  Therefore, the dynamics of the film thickness is coupled to the dynamics of the particle concentration.  The particles are assumed to be neutrally buoyant, i.e.\ of the same density as the liquid.  The pool is assumed to be large enough that the particle volume fraction concentration remains constant at $\phi_p$.  We ignore variations in particle concentration across the thickness of the suspension film, and assume $\phi$ to be independent of $y$.
	
	Though our intention is to address the observations of \segall\ in rotating horizontal cylinders partially filled with suspension, we choose to study the problem of a flat plate rising from a pool, as it simplifies the analysis.  While the equations describing the dynamics of the film in a curved wall which joins the suspension pool at the far end are almost identical to the ones we derive here, their numerical solution is a greater challenge.  It is clear, however, that the two problems are qualitatively similar, and are identical in the limit of the cylinder radius becoming very large compared to the film thickness, provided the fraction of cylinder filled with suspension is not small.

	From earlier investigations on coating flows, \cite{landau_levich42,middleman77,white_tallmadge65,spiers_etal74}, it is clear that the liquid film coating the plate can be divided into three regions (see Fig.~\ref{fig-schematic2}): (a) the static region at the bottom, where there is a balance between surface tension and gravity, (b) the ``entrainment'' region where surface tension, gravity and the viscous shear stress all play a role and (c) the region of constant film thickness at the top where gravity is balanced by the shear stress at the wall.  The solution of the problem is obtained by matching the solution of the entrainment region with that of the static region at one end and the straight film region at the other  \cite{landau_levich42,middleman77,white_tallmadge65,spiers_etal74}.   This solution yields, among other things, the asymptotic film thickness $w_c$.  It is in general a function of the Capillary number $\Ca \equiv v_w \eta/\sigma$; it varies as $\Ca^{1/6}$ when $\Ca \ll 1$, and becomes independent of it when $\Ca \gg 1$ \cite{middleman77}.

	In the static region, the balance of surface tension and gravity yields
\be
	\sigma \Frac{w''}{\left( 1 + w'^2 \right)^{3/2}} = \rho g \cos \theta,
\ee
the primes indicating differentiation with respect to $x$.  Integrating the above with the boundary condition $w' \to -\infty$ at $x=0$ \cite{levich62} gives
\be
	\Frac{w'}{\left( 1 + w'^2 \right)^{1/2}} =  \Frac{\rho g \cos \theta x^2}{2 \sigma} - 1
\ee
The point at which the slope vanishes is $x = (2\sigma/\rho g \cos \theta)^{1/2}$.  Following the prescription of \cite{landau_levich42}, we match the curvature at this point, 
\be
	w'' = \left( \Frac{2 \rho g \cos \theta}{\sigma} \right)^{1/2}
\label{eqn-matching}
\ee
with the curvature at the lower boundary of the entrainment region.

	In the region of constant film thickness, $w = w_c$, but its value is not known {\it a priori} -  every value of $w_c$ satisfies the balance between gravity and the wall shear stress, each corresponding to a particular value of the mass flux.  The value of $w_c$ must therefore be obtained by matching with the solution in the entrainment region.  Thus, the static and constant film thickness regions serve to set the boundary conditions for the entrainment region, and it is the solution in the entrainment region that determines the asymptotic film thickness $w_c$ \cite{landau_levich42,middleman77,white_tallmadge65,spiers_etal74}, and, as we shall see in what follows, the film dynamics.

	We now consider the entrainment region.  As stated earlier, we are interested not just in the steady state, but in obtaining a dynamical equation for $w$ that also allows variations in the $z$ direction. Following earlier studies on coating of a flat plate \cite{levich62,white_tallmadge65,spiers_etal74,middleman77}, we simplify the momentum balances by invoking the lubrication approximation, on the assumption that the film thickness $w$ is small compared to the distance in the $x$ direction over which there is variation of $w$.  The $x$- momentum balance for the suspension in the film reads
\be
  \eta_s \Frac{\partial^2 v_x}{\partial y^2} = \pder{p}{x} + \rho g \cos \theta.
\label{eqn-xmom}
\ee
Integrating the above with the no-slip boundary condition at the cylinder wall ($y=0$) and the stress free boundary condition at the liquid-air interface ($y=w$) gives
\be
  v_{x} = \Frac{1}{\eta_s} (\rho g \cos \theta + \pder{p}{x})  (\frac{y^2}{2} - yw) + v_w,
\label{eqn-vel_x1}
\ee
where $v_w$ is the velocity of the plate.  The continuity of normal stress across the interface yields the relation 
\be
	-p + \eta_s \pder{v_y}{y} - \sigma (\Frac{\partial^2 w}{\partial x^2} +  \Frac{\partial^2 w}{\partial z^2}) = p_{\mbox{\scriptsize atm}}	\quad \mbox{at $y=w$},
\label{eqn-ns_contin}
\ee
where we have again invoked the assumption that the gradient of w is small, i.e.\ $|\textfrac{\partial w}{\partial x}|, |\textfrac{\partial w}{\partial z}| \ll 1$, in linearizing 
the expression for the curvature of the interface.  The contribution of the viscous normal stress $\eta_s \textfrac{\partial v_y}{\partial y}$ was ignored by \cite{levich62,white_tallmadge65}, but retained by \cite{spiers_etal74}.  Retaining it improves the agreement of the predicted asymptotic coating thickness with experimental observations \cite{middleman77}.  We drop this contribution, thus considerably simplifying 
the analysis; if it is retained, the pressure cannot be eliminated from the equations of motion, and hence an additional field must be solved for.  Substituting \eqref{eqn-ns_contin} in \eqref{eqn-vel_x} with the above simplification yields
\be
  v_{x} = \left[ \Frac{\rho g \cos \theta}{\eta_s} - \Frac{\sigma}{\eta_s} \left( \Frac{\partial^3 w}{\partial x^3} + \Frac{\partial^3 w}{\partial z^2 \partial x} \right) \right] (\frac{y^2}{2} - yw) + v_w.
\label{eqn-vel_x}
\ee

	In the $z$- direction, flow is driven only by surface tension due to a gradient in the curvature.  The balance of momentum for this direction reads
\be
  \eta_s \frac{\partial^2 v_{z}}{\partial y^2} = -\sigma \left(\Frac{\partial^3 w}{\partial x^2 \partial z} +  \Frac{\partial^3 w}{\partial z^3} \right),
\ee
which upon integration, again using the no-slip boundary condition at the wall and no shear stress condition at the interface, yields
\be
  v_{z} = - \Frac{\sigma}{\eta_s} \left(\Frac{\partial^3 w}{\partial x^2 \partial z} +  \Frac{\partial^3 w}{\partial z^3} \right) (\Frac{y^2}{2} - yw).
\label{eqn-vel_z}
\ee

	The velocity $v_y$ is determined from the incompressibility condition,
\be
  \pder{v_y}{y} = -\pder{v_x}{x} - \pder{v_{z}}{z}.
\label{eqn-contin}
\ee
Upon substituting (\ref{eqn-vel_x}) and (\ref{eqn-vel_z}) in (\ref{eqn-contin}) and integrating, using the no-penetration boundary condition at the wall, we get
\bea
  v_y & = & \left[ \Frac{\rho g \cos \theta}{\eta_s} - \Frac{\sigma}{\eta_s} \left( \Frac{\partial^3 w}{\partial x^3} + \Frac{\partial^3 w}{\partial z^2 \partial x} \right) \right] \Frac{y^2}{2} \pder{w}{x} - \Frac{\sigma}{\eta_s} \left(\Frac{\partial^3 w}{\partial x^2 \partial z} +  \Frac{\partial^3 w}{\partial z^3} \right) \Frac{y^2}{2} \pder{w}{z} \\
& & + \left\{ \pder{}{x} \left[ \Frac{\sigma}{\eta_s} \left( \Frac{\partial^3 w}{\partial x^3} + \Frac{\partial^3 w}{\partial z^2 \partial x}\right) \right]  + \pder{}{z} \left[ \Frac{\sigma}{\eta_s} \left( \Frac{\partial^3 w}{\partial x^2 \partial z} + \Frac{\partial^3 w}{\partial z^3} \right) \right] \right\}
 (\Frac{y^3}{6} - \Frac{y^2w}{2})\
\label{eqn-vel_y}
\eea

	Finally, the kinematic constraint on the interface that its velocity be equal to that of the suspension adjacent to it yields
\be
 \pder{w}{t} + v_x \pder{w}{x} + v_z \pder{w}{z} = v_y \quad \mbox{at $y=w$}.
\label{eqn-kinem}
\ee
Substituting (\ref{eqn-vel_x}), (\ref{eqn-vel_z}) and
(\ref{eqn-vel_y}) in (\ref{eqn-kinem}), we have
\be
\pder{w}{t} + v_w \pder{w}{x} = \Frac{\partial}{\partial x} \left\{ \left[
\rho g \cos \theta - \sigma \left( \Frac{\partial^3 w}{\partial z^2 \partial x} + \Frac{\partial^3 w}{\partial x^3} \right) \right]
\Frac{w^3}{3 \eta_s} \right\} - \Frac{\partial}{\partial z}
\left[\Frac{\sigma w^3}{3 \eta_s} \left(\Frac{\partial^3 w}{\partial z^3} + \Frac{\partial^3 w}{\partial x^2 \partial z}\right)
\right] \label{eqn-w_dyn}
\ee
This is the equation for the dynamics of the film thickness - it is coupled to the particle volume fraction $\phi$ through the suspension viscosity $\eta_s(\phi)$.

	Eq.~(4) of \old\ can be recovered if we average \eqref{eqn-w_dyn} over the length of the entrainment region; neglecting the surface-tension terms for the moment and integrating \eqref{eqn-w_dyn} from the pool ($x=0$) to a distance $L$, we obtain
\be
	\pder{\ob{w}}{t} -\alpha_1 v_w = - \alpha_2 \Frac{\rho g}{\eta_s} \ob{w}^2
\label{eqn-eqn4}
\ee
where we have used
\bea
\Frac{w(x\!\!=\!\!0) - w(x\!\!=\!\!L)}{L} \approx \Frac{w(x\!\!=\!\!0)}{L} = \alpha_1, \nonumber \\
\Frac{[\cos\theta w^3/3 \eta_s]_{x=0} -  [\cos\theta w^3/3 \eta_s]_{x=L}}{L} \approx \Frac{[\cos\theta w^3/3 \eta_s]_{x=0}}{L} = \alpha_2 \ob{w}^2 /\ob{\eta_s}. \nonumber
\eea
As the film thickness rapidly decays to its asymptotic constant value (see Figs~\ref{fig-base} and \ref{fig-eig1}), it is sufficient to take $L$ to be a few multiples of $w(x\!\!=\!\!0)$.  Eq.~\eqref{eqn-eqn4} is {\em identical} to Eq.~(4) of our earlier paper \old\ if the surface tension term in the latter is  dropped. 
While the factors $\alpha_1$ and $\alpha_2$ are in general functions of $z$ and $t$, they were treated as constants in \old\ as a first approximation: this is formally correct if we assume that the shape of the $w(x)$ profile does not vary with $z$, even if its average does.  As the most striking feature in the observations of \cite{tirumkudulu_etal99,tirumkudulu_etal00} was an axial variation in the particle concentration and the film thickness, this is a reasonable first approximation.  Moreover, the results of this approximation are borne out by the full two-dimensional analysis in this work, as shown below.

Surface tension was included incorrectly 
in \old, making the naive guess that it entered simply at next order in gradients as $\sigma w \textfrac{\partial^2 w}{\partial z^2}$.  It is clear from \eqref{eqn-w_dyn} that it actually enters as $\textfrac{\sigma w^3}{3} \,\textfrac{\partial^4w}{\partial z^4}$, as acknowledged in our rebuttal to \acrcom.   Nevertheless, the physical basis for the dynamical equation for the film thickness in  \old\  is quite simple: it clearly follows from a balance of gravity, shear forces and surface tension, and can hardly be described  as ``having no basis whatsoever'' \acrcom.

	The equation governing $\phi$ is the mass balance for particles,
\be
\pder{\phi}{t} + \te{v} \te{\cdot} \nabla \phi =  -\nabla \te{\cdot} \te{j}
\ee
where $\te{j}$ is the flux of particles due to shear-induced diffusion \cite{leighton_acrivos87,phillips_etal92}.  The phenomena of shear-induced diffusion and migration have been addressed in several papers, and therefore does not require much description here.  The shear-induced flux has two parts, one driven by a gradient in the concentration and the other by a gradient in the shear rate, 
\be
	\te{j} = - f_c a^2 \gamdot \nabla \phi - f_s a^2 \phi \nabla \gamdot.
\label{eqn_particle_flux}
\ee
where $\gamdot$ is a measure of the shear rate, $a$ is the particle size.  The above can also be derived by treating the particles and fluid as separate, but interacting, phases \cite{nott_brady94}; combining the balances of the particle phase mass and momentum yields an equation identical to \eqref{eqn_particle_flux} to leading order in the gradients.  As the flux depends only on the magnitude of the shear rate and not its direction, $\gamdot$ must be a suitably chosen measure.  If $\te{S}$ is the rate of deformation tensor, the second invariant $\sqrt{\te{2 S\!:\!S}}$ is the natural choice for $\gamdot$ \cite{mauri_papageorgiou02} in two dimensional flows.  As we use the lubrication approximation, the only contributions to the shear rate rate come from $S_{xy} = S_{yx} = \textfrac{\partial v_x}{\partial y}$.  Further, since we assume no variation of the concentration in the $y$ direction (across the thickness of the film), we use thickness-averaged shear-rate and velocity,
\be
\ob{\gamdot} = 1/w \int_0^w \gamdot \, dy, \quad \ob{\te{v}} = 1/w \int_0^w \te{v} \, dy, 
\ee
 in the mass balance.  The particle mass balance then takes the form
\bea
\pder{\phi}{t} + \ob{v}_x \pder{\phi}{x} + \ob{v}_z \pder{\phi}{z} & = & \pder{}{x} \left[ f_c\, a^2 \, \ob{\gamdot} \, \pder{\phi}{x} + f_s \, a^2\, \phi \, \pder{\ob{\gamdot}}{x}
\right] \nonumber \\
& & + \pder{}{z} \left[ f_c \, a^2 \, \ob{\gamdot} \,
\pder{\phi}{z} + f_s \, a^2\, \phi \, \pder{\ob{\gamdot}}{z}
\right].
\label{eqn-phi_dyn}
\eea
Equations \eqref{eqn-w_dyn} and \eqref{eqn-phi_dyn} together describe the dynamics of the suspension film.

	The parameters $f_c$ and $f_s$ are in general direction dependent - the diffusivity in the velocity gradient direction generally differs from that in the vorticity direction.  However, we have taken them to be independent of direction in \eqref{eqn_particle_flux} for simplicity.  They are functions of the particle concentration.  The function $f_c(\phi)$ has been determined by \cite{acrivos_etal93} for migration in the vorticity direction in simple shear, but to our knowledge no direct measurement of $f_s$ exists.  Using the suspension balance model \cite{nott_brady94}, the diffusivities may be related to the particle phase pressure,
\be
	f_c = \chi \, 2/9 f(\phi) p'(\phi) \nonumber, \quad f_s = 2/9 f(\phi) p(\phi)/\phi
\label{eqn-fc_fs}
\ee
where $f(\phi)$ is the hindered settling function in the sedimentation of particles in a suspension and $p(\phi) \equiv \Pi/(\eta \gamdot)$ is the dimensionless pressure function, $\eta$ being the viscosity of pure fluid.  The hindered settling function is taken to be $f(\phi) = (1 - \phi)^{-5}$ \cite{russel_etal89}, and recent study of \cite{zarraga_etal00} gives $p(\phi) = \eta_s(\phi) \alpha(\phi)$ with $\alpha(\phi) =  1.89 \phi^3 e^{2.34 \phi}$.  The suspension relative viscosity is taken as $\eta_s = (1 - \phi/0.62)^{-1.83}$.

We must emphasize that the above relations for $f_c$ and $f_s$ are only rough estimates, as there is some uncertainty in the data of \cite{zarraga_etal00}; for instance, they have combined measurements of the stress in the suspension with an estimate of a normal stress in the particle phase to determine all the normal stresses.  Due to the uncertaintly in their dependence on $\phi$, we treat $f_c$ and $f_s$ as constants in our model, equal to their respective values at $\phi=\phi_p$.  The factor $\chi$ in the relation for $f_c$ is nominally equal to unity, but we have varied it to test the sensitivity of our predictions to the assumed forms above.

\subsection{Boundary conditions}

	The solution in the entrainment region is matched with that of the static region when the film thickness in the former becomes very large \cite{landau_levich42,levich62,middleman77}.  The matching is achieved by enforcing the boundary condition of \cite{landau_levich42} that the curvature $w''$ of the film must approach that given in \eqref{eqn-matching} at $w \to \infty$.  In the solution of \cite{landau_levich42}, the effect of gravity is neglected and the asymptotic behaviour for large $w$ is indeed such that $w''$ attains a constant value.  However, an inspection of  \eqref{eqn-w_dyn} shows that $w''' \to 1/\Sigma$ as $w \to \infty$, which gives a second derivative that varies linearly with $x$ as $w$ becomes large\cite{white_tallmadge65,spiers_etal74,middleman77}.  An approximate method has been used to satisfy the matching condition in some studies \cite{white_tallmadge65,spiers_etal74}, but we use the following simple condition that has the same qualitative effect: we would like $w'''$ to decay and $w''$ to asymptotically approach the constant value given in \eqref{eqn-matching} as $w \to \infty$, but since that isn't possible, we impose \eqref{eqn-matching} at the point where $w''' = 0$.  In other words, we {\em patch} the entrainment solution to the static solution where the curvature in the former is maximum.   Consequently, the boundary conditions in dimensionless form at the lower boundary of the entrainment region are
\be
	w'' = \left( \Frac{2 \rho g \cos \theta}{\sigma} \right)^{1/2}, \quad w''' = 0 \quad \mbox{at $x=0$}.
\label{eqn-bc1}
\ee

	At the upper end of the entrainment region, the solution is matched with that in the  constant film thickness region, and consequently the boundary conditions are
\be
	w' = 0, w'' = 0, w''' = 0 \quad \mbox{at $x \to \infty$}.
\label{eqn-bc2}
\ee

	For the particle concentration, we impose the condition that the concentration at the lower boundary is equal to that in the pool, i.e.
\be
	\phi = \phi_p \quad \mbox{at $x=0$}
\label{eqn-bc3}
\ee
and that the concentration gradient vanishes far from the pool,
\be
	\phi' = 0 \quad \mbox{at $x \to \infty$} 
\label{eqn-bc4}
\ee

\subsection{Base state solution}
\label{subsec-base}

	The base state, whose stability we determine in the following section, is the steady state
with no $z$- variation.  Under these conditions, equations \eqref{eqn-w_dyn} and \eqref{eqn-phi_dyn} reduce to 
\bea
v_w \pder{w}{x} & = & \Frac{d}{dx} \left[ \left( 
\rho g \cos \theta - \sigma \Frac{d^3 w}{d x^3} \right)
\Frac{w^3}{3 \eta_s}  \right],
\label{eqn-w_ss1}\\
\ob{v_x} \der{\phi}{x}  & = & \der{}{x} \left[ f_c \, a^2 \, \ob{\gamdot} \, \der{\phi}{x} + f_s \, a^2\, \phi \, \der{\ob{\gamdot}}{x}
\right] 
\label{eqn-phi_ss1}
\eea
We now define the following dimensionless variables,
\bea
	& w* = w/w_s, \quad x* = x/w_s, \quad z* = z/w_s, \quad \tau = v_w t /w_s & \nonumber \\
	& \te{v}* = \te{v}/v_w, \quad \gamdot* = \gamdot w_s/v_w, \quad \eta* = \eta_s(\phi)/\eta_s(\phi_p) & \nonumber
\eea
and the dimensionless parameters
\[
	\epsilon \equiv {a^2 \over w_s^2}, \quad \Sigma\equiv {\sigma \over 3 v_w\eta_s(\phi_p)},
\]
where $w_s \equiv \left(\Frac{3 v_w\eta_s(\phi_p)}{\rho g \cos \theta}\right)^{1/2}$ is the scale for the film thickness.  Henceforth, we work only in terms of these dimensionless variables, but drop the asterisks for the sake of convenience.  Equations \eqref{eqn-w_ss1}-\eqref{eqn-phi_ss1} then transform to
\bea
\Frac{(3 w^2 w' - w^3 {\cal N} \phi')}{\eta} \left(1 - \Sigma w''' \right)
- \Sigma w'''' \Frac{w^3}{\eta} - w' & = & 0,
\label{eqn-w_ss2}\\
\epsilon \left( f_c \, \ob{\gamdot}' \, \phi' + f_c \, \ob{\gamdot} \, \phi'' + f_s \, \phi' \, \ob{\gamdot}' + f_s \, \phi \, \ob{\gamdot}'' \right) - \left[ 1 - \left(1 - \Sigma w''' \right) \Frac{w^2}{\eta} \right] \phi'  & = & 0,
\label{eqn-phi_ss2}
\eea
where  the primes denote differentiation with respect to $x$, and ${\cal N} \equiv d \ln \eta/d\phi$.  The shear rate is $|dv_x/dy|$ averaged over the film thickness, giving $\ob{\gamdot} = (1 - \Sigma w''') 3 w/2$.  In dimensionless form, the boundary conditions \eqref{eqn-bc1} transforms to
\be
	w'' = (2/\Sigma)^{1/2}, \quad w''' = 0  \quad \mbox{at $x=0$}.
\ee
and the others remain unaltered from \eqref{eqn-bc2} - \eqref{eqn-bc4}.

From \eqref{eqn-w_ss2} we recover the equation given by \cite{middleman77,white_tallmadge65} for a homogeneous fluid by setting $\phi'$ to zero; further, if the gravitational body force is discarded, we recover the equation given by \cite{landau_levich42}.

\begin{figure}
\begin{center}
\begin{minipage}{3in}
\includegraphics[scale=0.45]{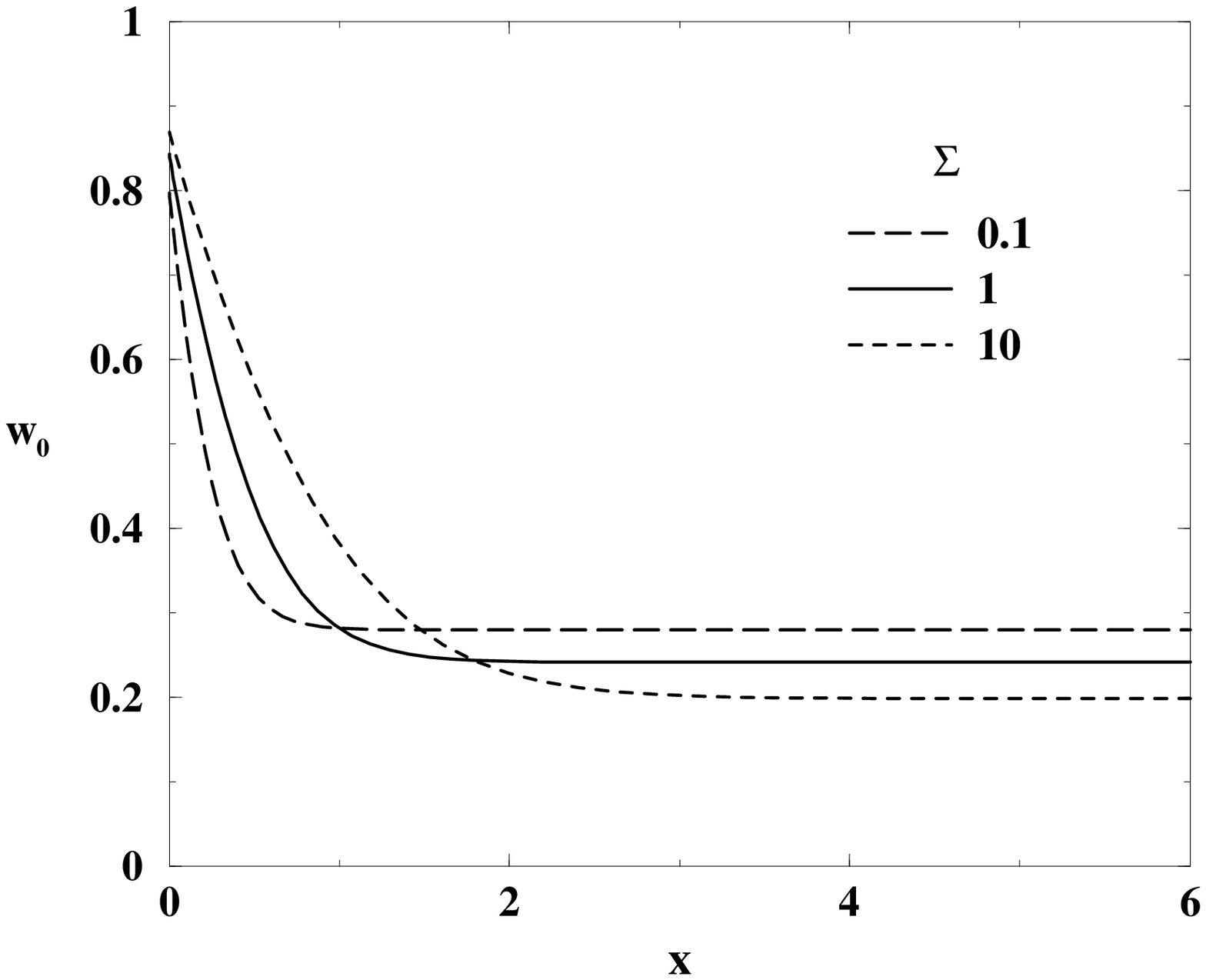} \\ \vspace*{-0.5em} (a)
\end{minipage}
\hspace*{2ex}
\begin{minipage}{3in}
\includegraphics[scale=0.45]{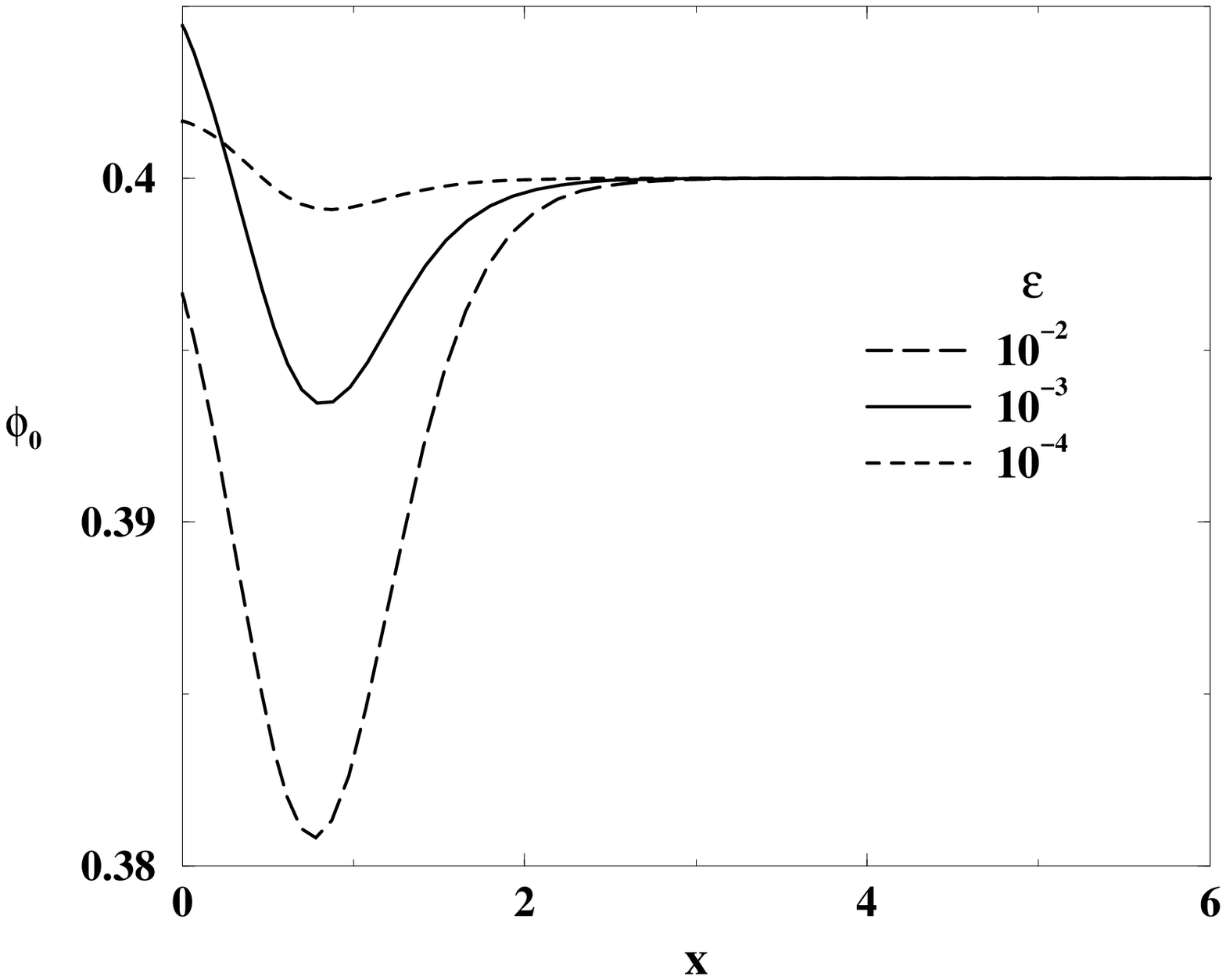}\\ \vspace*{-0.75em} (b)
\end{minipage}
\caption{(a) The base state film thickness for three values of the dimensionless surface tension. (b) The base state particle concentration profile for three values of $\epsilon$ with $\Sigma=1$.  The film thickness is insensitive to $\epsilon$ in the range studied.  In (a) $\epsilon=10^{-2}$ and in (b) $\Sigma=1$, and $\chi=1$ in both.\label{fig-base}}
\end{center}
\end{figure}

	The solution for the base state is obtained by integrating \eqref{eqn-w_ss2}-\eqref{eqn-phi_ss2} numerically from $x=0$ to a suitably large distance $x=x_\infty$.  It is easily seen that the thickness and concentration vary as $w(x) = w_c + A e^{-kx}$, $\phi(x) = \phi_c + r A e^{-kx}$ as $x \to \infty$, $w_c$ and $\phi_c$ being the asymptotic film thickness and particle concentration, respectively. As a result, one may impose, at $x = x_\infty$, 
the forms 
\bea
\label{eqn-alpheq}
	& w = w_c + \alpha, \quad w' = -k \alpha, \quad w'' = k^2 \alpha, \quad w''' = -k^3 \alpha, & \\
	& \phi = \phi_c + r \alpha, \quad \phi' = -k r \alpha. &\nonumber
\eea
The values of $k$ and $r$ are determined by substituting \eqref{eqn-alpheq} in 
\eqref{eqn-w_ss2}-\eqref{eqn-phi_ss2} and linearizing about $\alpha=0$.  Upon eliminating $r$ from the two equations, we get a quintic equation for $k$, and only one of its five roots is real and negative.  The solution was obtained by integrating from $x=x_\infty$ to $x=0$, with guesses for $w_c$ and $\phi_c$.  The guesses were refined until boundary conditions \eqref{eqn-bc1} and \eqref{eqn-bc3} are satisfied.  The solution is shown in Fig.~\ref{fig-base} for a few values of $\Sigma$ and $\epsilon$; it is clear that $w_0$ and $\phi_0$  almost equal their asymptotic values a short distance from the suspension pool.

\section{Linear Stability}
\label{sec-linear_stab}

	We now allow perturbations to the base state so that the film thickness and particle concentration have the form
\be
	w = w_{0}(x) + \tilde{w}(x,z,t), \quad \phi = \phi_0(x) + \tilde{\phi}(x,z,t),
\ee
where $w_0$ and $\phi_0$ are the base state solutions obtained in \S\ref{subsec-base}.  Assuming the perturbations are small, i.e.\ $\tilde{w} \ll w_0$, $\tilde{\phi} \ll \phi_0$, 
\eqref{eqn-w_dyn} and \eqref{eqn-phi_dyn} may be linearized about the base state.  Finally, taking the plate to be infinitely long in the $z$ direction, we may assume the solution to be in the form of normal modes in the $z$ direction, $\tilde{w} = \hat{w}(x) \, e^{i k_z z + s t}, \; \tilde{\phi} = \hat{\phi}(x) \, e^{i k_z z + s t}$.  We then get for the linearized equations for the perturbations
\bea 
s \hat{w} & = & \left[ \Frac{3c}{w_0^2}w_0' - \Sigma w_0^3(D^2-k_z^2)^2 -
3 w_0^2w_0'\Sigma (D^2-k_z^2)D+ \left(2 - \Frac{3c}{w_0} \right) D \right] \hat{w} 
\label{eqn-w_lin} \\
&-& {\cal N} \left[ w_0' + (w_0-c) D \right] \hat{\phi} \nonumber \\
s \hat{\phi} & = & -  \left[D - (1 - \Sigma w_0''') w_0^2 \left( D - {\cal N} \phi_0' \right) \right] \hat{\phi}  \label{eqn-phi_lin} \\
&+& \phi_0' \left[ 2 (1 - \Sigma w_0''') w_0  - \Sigma w_0^2 D^3 + \Sigma w_0^2 k_z^2 D\right] \hat{w} \nonumber \\
& + & \epsilon D
\left[ f_c \ob{\gamdot}_0 D \hat{\phi} + f_c \phi_0' \hat{\gamdot}
+ f_s \phi_0 D \hat{\gamdot} + f_s \ob{\gamdot}_0' \hat{\phi}  \right] 
 -  \epsilon \left[ k_z^2 f_c \ob{\gamdot}_0 \hat{\phi} -  k_z^2 f_s \phi_0 \hat{\gamdot} \right] , \nonumber 
\eea
with the shear rate perturbation $\hat{\gamdot}$ given by
\be
	\hat{\gamdot} = (1 - \Sigma w_0''') \Frac{3 \hat{w}}{2} - \Sigma \Frac{3 w_0}{2} \left(D^3 - k_z^2 D \right) \hat{w}  - (1 - \Sigma w_0''') \Frac{3 w_0}{2} {\cal N}  \hat{\phi}.
\ee
where $D$ represents the operator $\textfrac{\partial}{\partial x}$, and $c \equiv w_c - w_c^3$.

	Boundary conditions for \eqref{eqn-w_lin} and \eqref{eqn-phi_lin} are obtained by linearizing
\eqref{eqn-bc1}-\eqref{eqn-bc4},
\bea
	& D^2 \hat{w} = D^3 \hat{w} = \hat{\phi} = 0 \quad \mbox{at $x = 0$} &  \label{eqn-bc1_lin}\\
	& D \hat{w} = D^2 \hat{w} =  D^3 \hat{w} = D\hat{\phi} = 0 \quad \mbox{at $x =x_\infty$}.& \label{eqn-bc2_lin}
\eea

	Before solving the eigenvalue problem \eqref{eqn-w_lin}-\eqref{eqn-bc2_lin}, it is 
instructive to consider 
the simplest case of a film of uniform thickness dragged by (or draining from) 
a flat plate, as is the situation far above the pool.
As this problem is translation-invariant in $x$, we may Fourier-transform in the $x$ direction as well as $z$, 
$\tilde{w} = \hat{w} \, e^{i k_x x + i k_z z + st}$, $\hat{\phi} = \hat{\phi} \, e^{i k_x x + i k_z z + st}$.  The linearized equations describing this simplified problem are
\bea 
s \hat{w} & = & \left\{  - \Sigma w_0^3(k_x^2+k_z^2)^2 + i \left(-1 + 3 w_0^2 \right) k_x \right\} \hat{w} - \left\{ i {\cal N} w_0^3 k_x \right\} \hat{\phi} \label{eqn-w_lin2} \\
s \hat{\phi} & = & \left\{ 2 w_0  + i \Sigma w_0^2 k_x (k_x^2 +  k_z^2) -  \epsilon (k_x ^2 + k_z^2) \Frac{3 f_s \phi_0}{2}  \left[ 1 + i \Sigma w_0 k_x \left(k_x^2 + k_z^2 \right) \right] \right\} \hat{w}  \label{eqn-phi_lin2} \\
& - &  \left\{i ( 1 - w_0^2 )k_x  + \epsilon  (k_x^2 + k_z^2) \Frac{3 w_0}{2} \left(f_c - {\cal N}\phi_0 f_s \right) \right\} \hat{\phi}  \nonumber 
\eea
The eigenvalues of the above equations were determined and expanded as a series in $k_x$ and $k_z$ using the {\small MATHEMATICA} software package.  The results up to second order in the wave numbers are
\bea
	s & = & \mp \, {\cal N}^{1/2} w_0^2 \, k_x^{1/2} \pm \Frac{w_0^2}{4 \, {\cal N}^{1/2}} \, k_x^{3/2} - \frac{3}{4}\epsilon \, w_0  \left(f_c - {\cal N}\phi_0 f_s \right) \, (k_x^2 + k_z^2) \nonumber \\
 	& & +  i \left[ \pm \, {\cal N}^{1/2} w_0^2 \, k_x^{1/2} + (-1 + 2 w_0^2)\, k_x \pm \Frac{w_0^2}{4 \, {\cal N}^{1/2}} \, k_x^{3/2} \right].
\label{eqn-dispnrelnstraight}
\eea
When $k_x=0$, the result is essentially what was given in \old: 
there is an instability when the discriminant $E \equiv f_c - {\cal N}\phi_0 f_s$ 
turns negative, with the instability growth rate varying as $-E k_z^2$.  The 
mechanism of this instability, as given in \old, is the following:
a rise in $\phi$ at some axial position causes an increase in the 
local viscosity $\eta_s(\phi)$, leading to a fall in the local shear-rate there, 
in part by increasing the local film thickness $w$. Shear-induced diffusion causes 
particles to migrate towards regions of lower $\ob{\gamdot}$, further increasing 
$\phi$ in the region and therefore driving an instability.  We note here that \cite{mauri_papageorgiou02} analyzed the linear stability of wall-bounded plane Couette
flow, and predicted that the system is stable if driven at constant shear rate, but unstable at
constant shear stress provided $-f_c/7 < E < 0$; the condition of instability for
a uniform film is clearly less stringent.

There is also an instability at non-zero $k_x$, which varies as $k_x^{1/2}$ 
for small $k_x$, and is present even when $k_z=0$.  This ``pile-up'' 
instability, is not driven by shear-induced migration and hence does not involve the
discriminant $E$; it arises from the concentration 
dependence of the viscosity and the advection of concentration perturbations in the 
$x$ direction.  The stabilizing effect of surface 
tension comes in at fourth order in the wavenumbers in all directions.  

When the region near the pool is considered, the problem is not translation-invariant in
$x$, and we therefore return to the full problem \eqref{eqn-w_lin}-\eqref{eqn-bc2_lin}.
The eigenvalue problem is solved numerically by a Chebyshev collocation 
spectral method.  The physical domain $x= (0,x_\infty)$ is mapped to the computational 
domain $\xi = (-1,1)$ by the transformation
\begin{equation}
x= l \Frac{1-\xi}{1 + 2 l/x_\infty + \xi} \label{eqn-transformation}.
\end{equation}
The linearized equations are enforced at the $N$ collocation points, 
which are the roots of the $N^{\mbox{\scriptsize th}}$ order Chebyshev polynomial 
$T_N(\xi) \equiv \cos (N \cos^{-1} \xi)$.  The parameter $l$ is used to increase 
the density of collocation points near the pool, where there is rapid variation 
of the variables.  The variables $\hat{w}$ and $\hat{\phi}$ are expanded in 
Chebyshev polynomials, $\hat{w} = \sum_{i=1}^N a_i T_i(\xi)$, 
$\hat{\phi} = \sum_{i=1}^N b_i T_i(\xi)$, reducing 
\eqref{eqn-w_lin}-\eqref{eqn-bc2_lin} to a set of $2N$ linear algebraic 
equations for the $a_i$ and $b_i$.  The eigenvalues of the coefficient matrix 
give the growth rates $s_r$ of the perturbations. 
The real part of the leading eigenvalue $s_r^l$ is 
termed the leading growth rate, and its variation with $k_z$ and the dimensionless 
parameters in the problem are presented in Figs~\ref{fig-dispnrelnsigma} - \ref{fig-dispnrelnfc} below.  For all our computations, $x_{\infty}$ was at least 16, $l=10$ and $N=81$, which yielded eigenvalues with accuracy greater 
than $10^{-5}$.  The results presented below are for $\phi_p=0.4$; our results for lower $\phi_p$ are qualitatively the same, but the numerical accuracy of the eigenvalues is reduced.

\begin{figure}
\begin{center}
\psfrag{growth rate}[b][c]{$s_{r}^{l}$}
\psfrag{k}[c][]{$k_z$}
\psfrag{S}[c][]{$\Sigma$}
\includegraphics[scale=0.5]{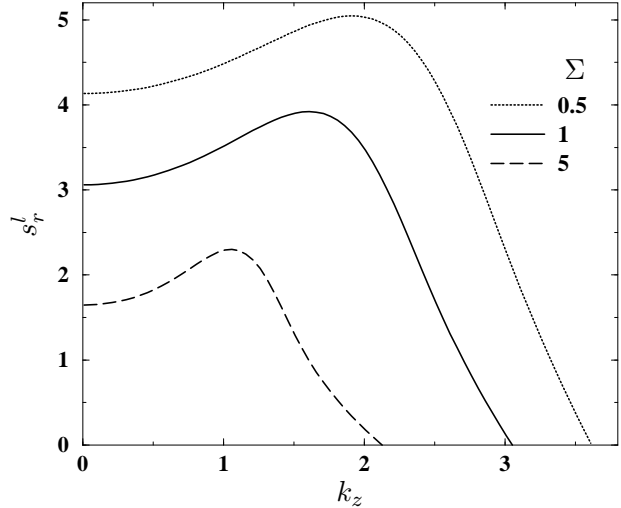}
\caption{The leading growth rate $s^l_r$ as a function of the wavenumber $k_z$ for some representative values of the dimensionless surface tension $\Sigma$.  Here the particle concentration in the pool is $\phi_p = 0.4$ and the parameter $\chi$ is 0.5, 
yielding a value of 4.47 for discriminant $E$, and $\epsilon=10^{-2}$.}
\label{fig-dispnrelnsigma}
\end{center}
\end{figure}

\begin{figure}
\begin{center}
\psfrag{growth rate}[][c]{$s_{r}^{l}$}
\psfrag{k}[c][]{$k_z$}
\includegraphics[scale=0.5]{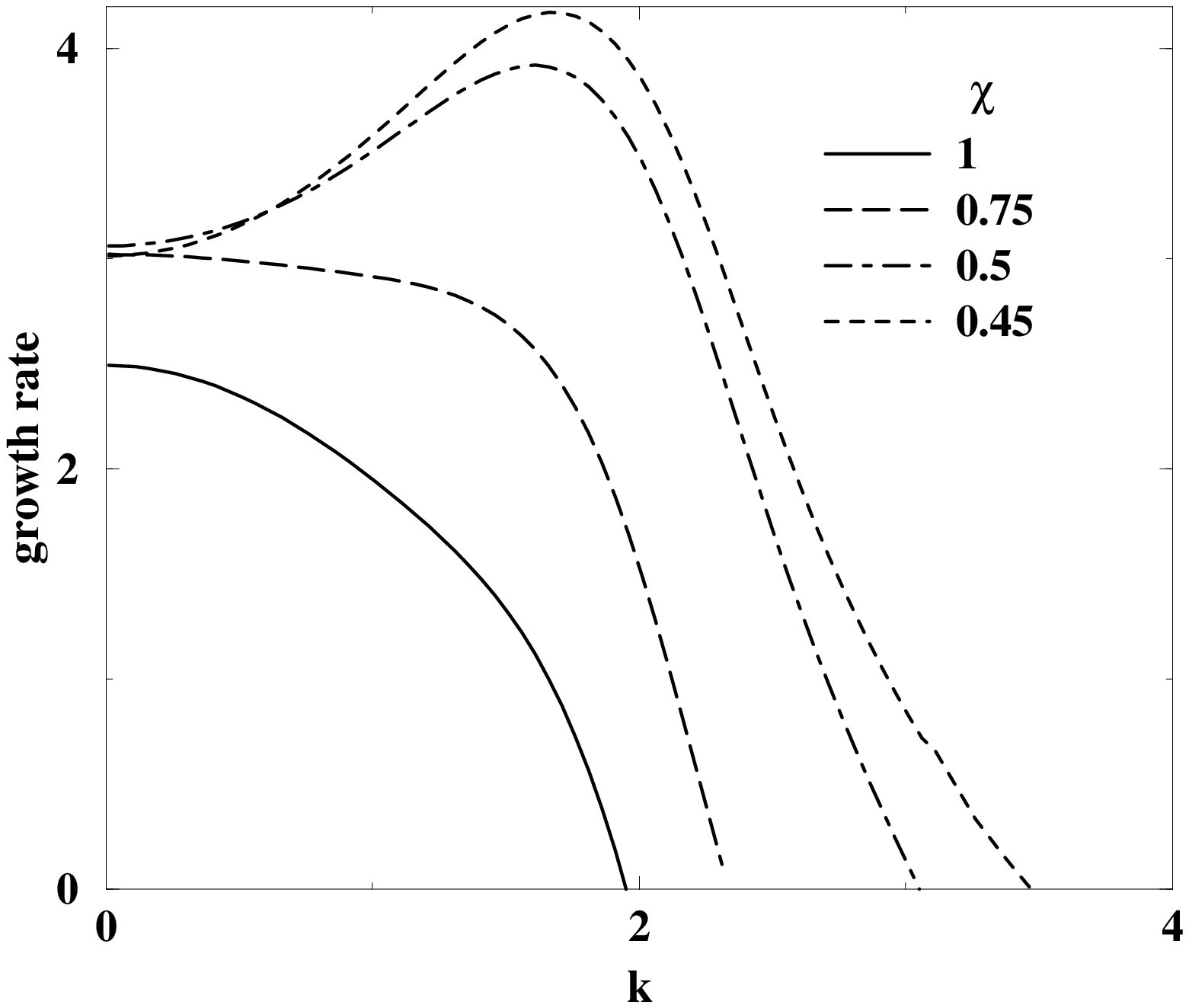}
\caption{The leading growth rate $s^l_r$ as a function of $k_z$ for different values of the factor $\chi$ (see Eq.~\ref{eqn-fc_fs}).  The 
values of the discriminant 
$E \equiv (f_c - {\cal N}\phi_0 f_s$) for the four cases are $57.76$, $31.11$, 
$4.47$ and $-0.86$ (in the order of decreasing $\chi$).  Parameter values are $\phi_p = 0.4$, $\Sigma=1$,  and $\epsilon=10^{-2}$.}
\label{fig-dispnrelnfc}
\end{center}
\end{figure}

\begin{figure}
\begin{center}
\psfrag{eigenfunction}[c][]{$\hat{w}(x)$, $\hat{\phi}(x)$}
\psfrag{f}[cm][]{$\hat{\phi}$}
\psfrag{w}[cm][]{$\hat{w}$}
\psfrag{x}[cm][]{$x$}
\includegraphics[scale=0.5]{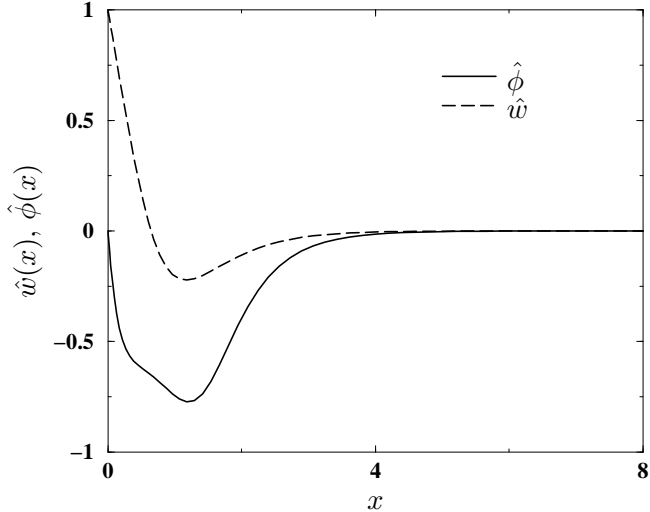}
\caption{The eigenfunctions for the particle concentration, $\hat{\phi}(x)$,  and the film 
thickness, $\hat{w}(x)$, for $k_z=1$. Notice that they are restricted to a fairly small range of $x$.  Parameter values are $\phi_p = 0.4$, $\Sigma=1$,  $\epsilon=10^{-2}$ and $\chi=1$.\label{fig-eig1}}
\end{center}
\end{figure}

	Figs~\ref{fig-dispnrelnsigma}  and \ref{fig-dispnrelnfc} 
show the dependence of $s^l_r$ on the parameters $\Sigma$ and 
$\chi$. 
Fig.~\ref{fig-eig1} shows the profile of the eigenfunctions for $\hat{\phi}$ and 
$\hat{w}$ as functions of $x$ for $k_z = 1$, showing that they are localised 
close to the pool. Several important points emerge from the figures: 
(i) Instability is always present, irrespective of the value of 
the dimensionless surface tension $\Sigma$. This is consistent with the discussion and 
results for the straight film \eqref{eqn-dispnrelnstraight}. 
(ii) The instability persists down to $k_z=0$ because, as seen in 
Fig.~\ref{fig-eig1}, variations along $x$ are always present in the eigenfunctions
so that the pile-up instability always arises;  
the mechanism for instabilities with appreciable variation along $z$ is, 
as we have said above, essentially that outlined in \old\ riding on 
the pile-up instability. If the discriminant $E$ is below a critical value $E_c$, 
the growth rate above the baseline
value at $k_z = 0$ varies as $k_z^2$ for small $k_z$, as 
stated in \cite{govindarajan_etal01}.  While the one-dimensional analysis of
\old\ predicted $E_c=0$, our full two-dimensional here predicts
$E_c > 0$ (see Fig.~\ref{fig-dispnrelnfc}).
(iii) Surface tension tends to stabilize modes at large wavenumber and hence determines 
the range of unstable wavenumbers. This can be seen clearly in 
Fig.~\ref{fig-dispnrelnsigma}.   
(iv) Changing the value of the discriminant $E = (f_c - {\cal N} \phi_0 f_s)$ 
has the effect discussed in \old: a large positive 
value of that parameter, controlled here by the parameter $\chi$, greatly reduces 
the growth rate (which would in fact go negative if it were not for the 
effect of variation along $x$). This can be seen in Fig.~\ref{fig-dispnrelnfc},  
where the curve for $\chi = 0.45$ corresponds most closely to a case dominated 
by modes varying along $z$, and hence shows the increase in growth-rate 
with increasing wavenumber.   
(v) For typical parameter values considered, the most unstable 
wavenumber is at $k_z \approx 1$, which implies a wavelength of $\approx 2 \pi w_s$. 
For the experiments of \cite{tirumkudulu_etal00}, $w_s$ in the range 1-3 mm, 
and hence the wavelength of the fastest growing perturbations would be around 1 cm,  
which is within their observed range of the distance between adjacent bands.

\section{Summary and conclusion}
\label{concl}

	We have shown that a film of suspension coating a wall rising from a suspension
pool can display an instability to perturbations that cause segregation of particles in axial bands. 
This instability is particularly pronounced when the discriminant $E$,
which depends on the dimensionless shear-induced
diffusivities and the suspension viscosity, is below a critical value $E_c$.  This bears out the results we obtained in an earlier paper \old, now in a complete 
analysis in which the equation governing the dynamics of the film thickness is
derived from the Stokes equations and the kinematic condition at the free interface,
and a fully two-dimensional  stability analysis carried out.  The 
instability comes from the coupled dynamics of film thickness $w$
and particle concentration $\phi$.  The condition for instability we predict is
less stringent than the condition $-f_c/7 < E < 0$ for plane Couette flow at constant shear stress given by \cite{mauri_papageorgiou02}.

We observe two kinds of instabilities: the first is the
{\em axial} instability proposed in our earlier one-dimensional treatment, in which axial
variations of the film thickness and particle concentration are amplified with time. The instability growth rate for this instability varies as the square of the axial wavenumber $k_z$ for small $k_z$,
precisely as in the one-dimensional model of \old.  The mechanism of this instability is that
a rise in $\phi$ at some axial position causes an increase in the 
local viscosity $\eta_s(\phi)$, leading to a fall in the local shear-rate $\gamdot$, 
in part by increasing $w$; shear-induced diffusion causes 
particles to migrate towards regions of lower $\gamdot$, further increasing 
$\phi$ in the region and therefore driving an instability.  
In the second ``pile-up'' instability, variations along the $x$ direction (i.e. of wall motion) are amplified, and the growth rate is finite even at $k_z=0$.  This mechanism of this instability is that a rise in $\phi$ at one position results in a rise in the local viscosity and hence a reduction in the veolcity.  This reduces the advection flux of particles from that spot, causing a further accumulation of particles downstream.

 While we find that the eigenfunctions
always have a significant $x$-dependence, and therefore the instability is inherently two-dimensional, the dependence of the instability 
growth rate on the axial wavenumber is as in the one-dimensional model \old when $E < E_c$.
Surface tension plays a stabilizing role at larger wavenumbers, but in a manner 
different in detail from that proposed in \old. 
To see whether this mechanism is indeed responsible for the observations of 
\cite{tirumkudulu_etal99,tirumkudulu_etal00}, 
independent measurements of
the dimensionless shear-induced diffusivities $f_c$ and $f_s$ (cf. Eq.~\ref{eqn_particle_flux})
are required.

\section{Acknowledgements}
\label{ack}
For support, RG thanks the DR\&DO, PRN the DST, and SR the DST through the Centre for Condensed Matter Theory.  PRN gratefully acknowledges discussions with Professor~K. S. Gandhi on the coating flows of viscous liquids.

\newpage

\centerline{\sc \bf LIST OF FIGURE CAPTIONS}
\vspace*{2em}
Figure 1: Schematic diagram of film coating in a horizontal rotating cylinder.
\vspace*{1em}

Figure 2: Coating of a vertical plate rising with velocity $v_w$ from a pool of suspension.
\vspace*{1em}

Figure 3: The base state film thickness for three values of the dimensionless surface tension. (b) The base state particle concentration profile for three values of $\epsilon$ with $\Sigma=1$.  The film thickness is insensitive to $\epsilon$ in the range studied.  In (a) $\epsilon=10^{-2}$ and in (b) $\Sigma=1$, and $\chi=1$ in both.
\vspace*{1em}

Figure 4: The leading growth rate $s^l_r$ as a function of the wavenumber $k_z$ for some representative values of the dimensionless surface tension $\Sigma$.  Here the particle concentration in the pool is $\phi_p = 0.4$ and the parameter $\chi$ is 0.5, 
yielding a value of 4.47 for discriminant $E$, and $\epsilon=10^{-2}$.
\vspace*{1em}

Figure 5: The leading growth rate $s^l_r$ as a function of $k_z$ for different values of the factor $\chi$ (see Eq.~\ref{eqn-fc_fs}).  The 
values of the discriminant 
$E \equiv (f_c - {\cal N}\phi_0 f_s$) for the four cases are $57.76$, $31.11$, 
$4.47$ and $-0.86$ (in the order of decreasing $\chi$).  Parameter values are $\phi_p = 0.4$, $\Sigma=1$,  and $\epsilon=10^{-2}$.
\vspace*{1em}

Figure 6: The eigenfunctions for the particle concentration, $\hat{\phi}(x)$,  and the film 
thickness, $\hat{w}(x)$, for $k_z=1$. Notice that they are restricted to a fairly small range of $x$.  Parameter values are $\phi_p = 0.4$, $\Sigma=1$,  $\epsilon=10^{-2}$ and $\chi=1$.

\pagestyle{empty}

\newpage

\psfrag{w}{$w$}
\psfrag{r}{$r$}
\psfrag{v}[b]{$v_w$}
\psfrag{cr}[bc]{\rule{0.15em}{0em}$R$}
\psfrag{Suspension pool}[bl]{suspension pool}
\psfrag{film}[bl]{film}
\psfrag{q}[bl]{$\theta$}
\begin{center}
\includegraphics[scale=1]{schematic2.eps}\\
\vspace*{5em}
Figure 1
\end{center}

\newpage

\psfrag{vw}[b]{$v_w$}
\psfrag{w}{$w$}
\psfrag{x}{$x$}
\psfrag{y}{$y$}
\psfrag{z}{$z$}
\begin{center}
\includegraphics[scale=0.75]{schematic3.eps}\\
\vspace*{5em}
Figure 2
\end{center}

\newpage

\begin{center}
\includegraphics[scale=0.6]{w0.eps}\\
\vspace*{5em}
Figure 3a
\end{center}

\newpage

\begin{center}
\includegraphics[scale=0.6]{phi0.eps}\\
\vspace*{5em}
Figure 3b
\end{center}

\newpage

\begin{center}
\psfrag{growth rate}[b][c]{$s_{r}^{l}$}
\psfrag{k}[c][]{$k_z$}
\psfrag{S}[c][]{$\Sigma$}
\includegraphics[scale=0.6]{Sig_half.eps}\\
\vspace*{5em}
Figure 4
\end{center}

\newpage

\begin{center}
\psfrag{growth rate}[][c]{$s_{r}^{l}$}
\psfrag{k}[c][]{$k_z$}
\includegraphics[scale=0.6]{fcfac_c.eps}\\
\vspace*{5em}
Figure 5
\end{center}

\newpage

\begin{center}
\psfrag{eigenfunction}[c][]{$\hat{w}(x)$, $\hat{\phi}(x)$}
\psfrag{f}[cm][]{$\hat{\phi}$}
\psfrag{w}[cm][]{$\hat{w}$}
\psfrag{x}[cm][]{$x$}
\includegraphics[scale=0.6]{eig1.eps}\\
\vspace*{5em}
Figure 6
\end{center}

\end{document}